\def\be{\begin{equation}}
\def\te{\end{equation}}
\def\bea{\begin{eqnarray}}
\def\nn{\nonumber}
\def\tea{\end{eqnarray}}

\def\a{\alpha}
\def\b{\beta}

\def\d{\delta}

\def\f{\phi}

\def\k{\kappa}

\def\m{\mu}

\def\o{\omega}
\def\p{\pi}
\def\q{\theta}

\def\s{\sigma}

\def\D{\Delta}

\def\O{\Omega}

\newskip\humongous \humongous=0pt plus 1000pt minus 1000pt

\newif\ifdtup

\def\ha{{1\over 2}}
\def\hx{\hat x}
\def\hp{\hat p}

\documentstyle[12pt]{article}

\makeatletter                    
\@addtoreset{equation}{section}  
\makeatother                     

\begin{document}

\title{Squeezed Vacua and the
Quantum Statistics of Cosmological Particle Creation}

\author{B. L. Hu, G. Kang, A. Matacz
\thanks{Permanant Address: Department of Physics, University of Adelaide,
5005, Australia}\\
{\small Department of Physics, University of Maryland,
College Park, MD 20742, USA}\\
{\small (umdpp 93-163)}}

\maketitle
\begin{abstract}

We use the language of squeezed states to
give a systematic description of two issues in cosmological particle creation:
\\
a) Dependence of particle creation on the initial state specified. We consider
in particular the number state, the coherent and the squeezed state.
b) The relation of spontaneous and stimulated particle creation and their
dependence on the initial state.
We also present results for the fluctuations in particle number in
anticipation of its relevance to defining noise in quantum fields and
the vacuum susceptibility of spacetime.

\end{abstract}
\newpage

\section{Introduction}

Cosmological particle creation \cite{Par69,SexUrb,Zel70,ZelSta,Hu72}
is a physical process of basic
theoretical interest in quantum field theory in curved spacetime \cite{BirDav},
and important applied interest in the quantum dynamics of the early universe
\cite{ZelSta,HuPar,HarHu}.
In this paper we use the language of squeezed states \cite{sqst} to
give a systematic description of two interelated issues:\\
a) Dependence of particle creation on the initial state. We consider
in particular the number state, the coherent and the squeezed state.\\
b) The relation of spontaneous and stimulated particle creation and their
dependence on the initial state.\\
We also present the result for the fluctuations in particle number in
anticipation of its relevance to defining noise in quantum fields.\\

Both of these issues have been explored before, albeit in a restricted context.
The use of Bogolubov transformation relating the canonical operators
between the {\it in} and {\it out} states was introduced by
Parker \cite{Par69}
in cosmological particle creation. He also derived the evolutionary operator
based on earlier work of Kamefuchi and Umezawa \cite{KamUme} and briefly
discussed induced particle creation. Zel'dovich \cite{Zel70} first pointed
out that cosmological particle creation is the quantum version of parametric
amplification of classical waves.
The connection of cosmological particle creation
with processes in quantum optics was thus noticed
more than 20 years ago \cite{Zel70,ZelSta,Hu72,Hu74,Gri74,Ber75}.\\

Cosmological particle creation in coherent states was discussed by Hu
\cite{Hu72}
and Berger \cite{Ber75}. The former used a close analogy with a model in
quantum optics based on the quantum statistics of coupled harmonic oscillators
( e.g., \cite{Mollow,HuMat2}).
Entropy generation associated with thermal particle creation in an
exponentially expanding universe was discussed by Parker \cite{Par76}.
The distinction of the number state and the coherent state
(in the so-called $N$ and $P$ representation)  in the question of
entropy generation
in particle creation was first noticed by Hu and Pavon \cite{HuPav} in
their search for an intrinsic measure of field entropy.
They found that the variance of particle number 
is a monotonically increasing function
of time in the $P$ representation, and that an increase
of particles in the $N$ representation is related to the fact that one has
chosen an initial state which is an eigenstate of the number operator,
as is often the case in most discussions of cosmological particle creation.
The phase-particle number
uncertainty relation implies that this choice amounts to assuming an
initial state with random phase.
Kandrup \cite{Kan} has further clarified these
points. Entropy generation in particle creation with interactions was
investigated by Hu and Kandrup \cite{HuKan}, who discussed both spontaneous and
stimulated production of particles.\\

Since the concept of squeezed state was introduced to quantum optics in the
seventies \cite{sqst},
there has been much progress in its experimental realizations
and theoretical implications. The adoption of the language of squeezed states
to cosmological particle creation was introduced recently by
Grishchuk and Sidorov \cite{GriSid}.
Although the physics is not new  (this was also pointed out by Albrecht
{\it et al} \cite{Alb93} in the inflationary cosmology context)
and the results are largely known,
the use of rotation and squeeze operators give an alternative description
which allows one to explore new avenues based on interesting ideas
developed in quantum optics.
More recent work on entropy generation in cosmological perturbations
by Brandenberger and coworkers \cite{BPM} and  Gasperini and
Giovannini \cite{GasGio} make use of coarse-graining
via a random phase approximation using the language of squeezed states.
Matacz \cite{Mat93} has used the squeezed state formalism as a starting point
for the study of decoherence of cosmological inhomogeneities in the coherent-
state representation. \\

The issues of initial states and entropy generation have been discussed
in restricted conditions, and the issue of spontaneous and stimulated
production
has only been touched upon before. For the sake of completeness,
in this work we will address these issues under a common framework,
and present the results for different initial states
(the number state, the coherent state and the squeezed state) . In Sec. 2 we
give a short  summary of particle creation described both in the old
language of Bogolubov transformations and in the new language of squeezed
states, mainly to present the basic concepts and introduce the terminology.
Readers familiar with cosmological particle creation and the squeezed state
description can go directly to Sec. 3.
In Sec. 3 we give the result of spontaneous and stimulated production
for different initial states for bosons.
In Sec. 4 we work out the change in the fluctuations in particle number.
This is in anticipation of its relation to defining noise in
quantum fields and vacuum susceptibility in quantum processes
in curved spacetimes.
We aim in this simple paper only to show how the old and the new
language can be used together to describe the quantum statistics of
particle creation. Exploration
of the implications of these processes will be described with more
detail in later works.\\

\section{Particle Creation and Squeezed State}

\subsection{Particle Creation via Bogolubov Transformation}
Readers familiar with the process of cosmological particle creation
can skip this subsection. For the original work, see \cite{Par69}.
Our summary here is adopted from \cite{HuKan} with slight modifications
in the discussion of spontaneous versus stimulated creation.

Consider a massive ($m$) scalar field $\Phi$ coupled arbitrarily
($\xi$) with a background spacetime with metric $g_{\mu \nu}$ and
scalar curvature $R$.  Its dynamics is described by the Lagrangian
density
\be
 L(x) =  -\frac{1}{2} \sqrt{-g} [ g^{\mu \nu}(x)\nabla_\mu
\Phi \nabla_\nu \Phi -  \left[ m^2 + (1 - \xi) \frac{R}{6} \right] \Phi^2(x)].
\te   
Here $\xi = 0$ and 1 denotes, respectively, conformal and minimal
coupling.  The scalar field satisfies the wave equation
\begin{equation}
\left[ \Box + m^2 + (1 - \xi) \frac{R}{6}\right] \Phi (\vec x,t) = 0,
\end{equation}  
where $\Box = g^{\mu \nu} \nabla_\mu \nabla_\nu$ is the
Laplace-Beltrami operator defined on the background spacetime.

In the canonical quantization approach, one assumes a foliation of
spacetime into dynamically evolving, time-ordered, spacelike
hypersurfaces $\Sigma$, expands the field on $\Sigma$ in normal
modes, imposes canonical commutation relations on the time-dependent
expansion functions now regarded as creation and annihilation
operators, defines the vacuum state, and then constructs the Fock
space.  In flat space, Poincar\'{e} invariance guarantees the
existence of a unique global Killing vector $\partial_t$ orthogonal
to all constant-time spacelike hypersurfaces, an unambiguous
separation of the positive-and negative-frequency modes, and a unique
and well-defined vacuum.  In curved spacetime, general covariance
precludes any such privileged choice of time and slicing.  There is
no natural mode decomposition and no unique vacuum.  At any
constant-time slice, one can expand the field $\Phi$ in terms of a
complete set of (spatial) orthonormal modes $u_{\vec k}(\vec x)$ \cite{Par69}
\begin{equation}
\Phi (x) = \sum_{\vec k} [\psi_{\vec k}(t)u_{\vec k}(\vec x)
+ \psi^\dagger _{\vec k}(t)u^*_{\vec k}(\vec x)].
\end{equation}   
After second quantization, the fields $\Phi$ and their amplitudes
$\psi_{\vec k}$ become operator-valued functions.  Write
\begin{equation}
\psi_{\vec k}(t) = a_{\vec k}(t) \phi_{\vec k}(t),
\end{equation}  
where $a_{\vec k}$ are the annihilation operators and the (c-number)
functions $\phi_{\vec k}(t)$ obey the wave equation derived from (2.2).  The
canonical commutation rules on $\Phi$ imply these conditions on $a_{\vec k}$
and $a^\dagger_j$, i.e.,
\begin{equation}
[a_{\vec k},a_{\vec j}] = [a^\dagger_{\vec k}, a^\dagger_{\vec j}]=0 \; \;
{\rm and} \; \; [a_{\vec k},a^\dagger_{\vec j}]=\delta_{\vec k \vec j}.
\end{equation}  
Assume that initially
a vacuum state $\mid 0 \rangle$ at $t_0$ can be defined by
\begin{equation}
a_{\vec k} \mid 0 \rangle_{t_0} = 0,
\end{equation}  
and a Fock space can be constructed from the $n$-particle states by the
action of the creation operators.  At a later time, say $t_1=t_0 + \Delta t$,
the vacuum state defined at $t_0$ will no longer be vacuous,
since the annihilation operator $b_{\vec k}(t_1)$ at $t_1$ is not equal to
$a_{\vec k}(t_0)$.  In general, they are related by a set of Bogolubov
transformations
\begin{equation}
b_{\vec j}(t_1) = \sum_{\vec k} [\a_{\vec j \vec k}(t)a_{\vec k} +
\beta^*_{\vec j \vec k}(t)a^\dagger_{\vec k}].
\end{equation}  
A new vacuum state $\mid 0)$ at $t_1$ can be defined by
\begin{equation}
b_{\vec j} \mid 0)_{t_1}=0
\end{equation}  
and from this a new Fock space can be constructed.  One can easily see
that $a_{\vec k} \mid 0) \neq 0$.  The two vacua are different by the
coefficients $\alpha, \beta$ whose time dependence are determined by
the amplitude functions $\phi_{\vec k}(t)$.  In particular, any $\phi_{\vec k}$
with only a positive-frequency component initially at $t_0$ will
acquire a negative-frequency component at $t_1$.  The new vacuum at
$t_1$ now contains
\begin{equation}
s_{\vec j} = (0 \mid N_{\vec j} \mid 0)
= \sum_{\vec k} \mid \beta_{\vec j \vec k} \mid^2
\end{equation}   
particles, where
\begin{equation}
N_{\vec j} \equiv b^\dagger_{\vec j} b_{\vec j}
\end{equation}  
is the particle number operator.  From (2.7) one sees that
$\beta_{\vec j \vec k}$
measures the negative-frequency component generated by dynamics.  In
curved space the inequivalence of Fock representation due to the lack
of a global timelike Killing vector makes the constant separation of
positive-and negative-frequency modes in general impossible.  The
mixing of positive- and negative- frequency modes in second-quantized
form leads to vacuum particle creation.  Particle creation may arise
from topological, geometrical, or dynamical causes.  In cosmological
spacetimes the inequivalence of vacua appears at different times of
evolution, and thus cosmological particle creation is by nature a
dynamically induced effect.  Note that we are dealing here with a
free-field: particles are not produced from interactions,
but rather from the excitation (parametric amplification \cite{Zel70})
of vacuum
fluctuations by the changing background gravitational field.

\subsubsection {Spontaneous Production}

For spacetimes with certain symmetries, some natural mode
decomposition may present itself.  For example, in the class of
conformally static spacetimes (e.g., Robertson-Walker universe),
where the metric is conformally related to a static spacetime (e.g.,
the Minkowski metric),
\begin{equation}
g_{\mu \nu} (x) = a^2 (\eta)\eta_{\mu \nu},
\end{equation}  
where $a$ is the conformal factor, there exists a global conformal
Killing vector $\partial_\eta$, where $\eta = \int dt/a(t)$ is the
conformal time.  Thus the vacuum defined by the mode decomposition
with respect to $\partial_\eta$ is a globally well-defined one, known
as the conformal vacuum.  For conformally-invariant fields [e.g., a massless,
scalar field with $\xi = 0$ in (2.1)] in conformally-static spacetimes,
it is easy to see that there is no particle creation \cite{Par69}.
Thus any small
deviation from these conditions, e.g., small $m,\xi$, can be treated
perturbatively from these states.
Consider the spatially-flat Robertson-Walker metric with line element
\begin{equation}
ds^2 = a^2 (\eta)(d\eta^2 -d \vec x^2).
\end{equation}  
The scalar fields can be separated into modes
\begin{equation}
\Phi (\eta , \vec x) = \sum_{\vec k} \phi_{\vec k}(\eta)
e^{i \vec k \cdot \vec x},
\end{equation}  
where $\phi_{\vec k}$ are the amplitude functions of the $\vec k$th mode.
 Define
new field variables $a(\eta) \phi_{\vec k}(\eta)=\chi_{\vec k}(\eta)$.
 From the
wave equation (2.2) for the $\vec k$th mode $\chi_{\vec k}(\eta)$ satisfies
\begin{equation}
\chi^{\prime \prime}_{\vec k}
(\eta)+[k^2+(m^2-\xi R/6)a^2]\chi_{\vec k}(\eta)=0.
\end{equation}  
where $k \equiv |\vec k | $.
One sees that, for massless ($m=0$) conformally coupled ($\xi=0$)
fields, $\chi_{\vec k}$ admits solutions
\begin{equation}
\chi_{\vec k}(\eta)=Ae^{i\Omega_{\vec k} \eta}+Be^{-i\Omega_{\vec k} \eta},
\end{equation}  
which are of the same form as travelling waves in flat space.  Since
$\Omega_{\vec k} =k=$ const, the positive- and negative-frequency components
remain separated and there is no particle production \cite{Par69}.  More
generally, the wave equation for each mode has a time-dependent
natural frequency given by
\begin{equation}
\Omega_{\vec k}^2(\eta)=k^2 + (m^2 - \xi R/6) a^2
\equiv \omega_{\vec k}^2 a^2.
\end{equation}   
The negative-frequency modes can thus be excited by the dynamics of
the background through $a(\eta)$ and $R(\eta)= 6 a''/a^3 $ (a prime
denotes $d/d \eta$).  In analogy with the time-dependent
Schr\"{o}dinger equation, one can view the $(m^2-\xi R/6)a^2$ term in
(2.14) as a time-dependent potential $V(\eta)$ which can induce
backscattering of waves \cite{Zel70,Hu72}.
The number of created particles in the $\vec k$th
mode is given in terms of $\chi^\prime_{\vec k}$ and $\chi_{\vec k}$ by
\begin{equation}
s_{\vec k}=\mid \beta_{\vec k} \mid^2 = \frac{1}{2 \Omega_{\vec k}}
( \mid \chi^\prime_{\vec k} \mid^2
+ \Omega^2_{\vec k} \mid \chi_{\vec k} \mid^2) - \frac{1}{2}.
\end{equation}    
The energy density associated with these particles is given by the
expectation value of the 00 component of the conformal
energy-momentum tensor
with respect to the conformal vacuum:
\begin{eqnarray}
\rho _0 & = & \langle   0 \mid \Lambda^0_0 \mid 0 \rangle \nonumber \\
        & = & \frac{1}{a^4} \int \frac{d^3k}{2(2\pi)^3}
       (\mid \chi^\prime_{\vec k} \mid^2
      + \Omega^2_{\vec k} \mid \chi_{\vec k} \mid^2) \nonumber \\
       & = & \frac{1}{a^4} \int \frac{d^3k}{(2 \pi)^3} (2s_{\vec k} + 1)
           \frac{\Omega_{\vec k}}{2}.
\end{eqnarray}   
In a Hamiltonian description of the dynamics of
a finite system of parametric oscillators, the Hamiltonian is simply
\begin{equation}
H_0(t)=\frac{1}{2}\sum_{k}(\pi_{\vec k}^2 + \Omega_{\vec k}^2q_{\vec k}^2)
      =\sum_{k}(N_{\vec k}+\frac{1}{2})\Omega_{\vec k},
\end{equation}
Comparing this with (2.18) one can identify $\mid \chi_{\vec k} \mid^2$
and $\mid \chi^\prime_{\vec k} \mid^2$ with
the canonical coordinates $q_{\vec k}^2$ and moment $\pi^2_{\vec k}$,
the eigenvalue of $H_0$
being the energy $E_{\vec k}=(N_{\vec k} + \frac{1}{2}) \Omega_{\vec k}$.
The analogy of particle creation with parametric amplification is formally
 clear:
(2.17) defines the number operator
\begin{equation}
N_{\vec k} = \frac{1}{2 \Omega_{\vec k}} (\pi_{\vec k}^2 +
\Omega_{\vec k}^2 q_{\vec k}^2) - \frac{1}{2},
\end{equation}  
and (2.18) says that
the energy density of vacuum particle creation comes from the
amplification of vacuum fluctuations $\hbar \Omega_{\vec k}/2$ by the factor
${\cal A}_{\vec k}=2s_{\vec k}+1$.

\subsubsection {Stimulated Production}

Equation (2.18) gives the vacuum energy density of particles produced
from an initial vacuum, a pure state.  If the initial state at $t_0$
is a statistical mixture of pure states, each of which contains a
definite number of particles, then an additional mechanism of
particle creation enters.  This is categorically known as induced
creation.  In particular, as already pointed out in the original
paper of Parker \cite{Par69},
if the statistical density matrix $\mu$ is diagonal
in the representation whose basis consists of the eigenstates of the
number operators $a^\dagger_{\vec k}a_{\vec k}$ at time $t_0$
,then for bosons, this process increases the average
number of particles (in mode $\vec k$ in a unit volume) at a later time
$t_1$ over the initial amount:
\begin {eqnarray}
( N_{\vec k}(t) ) & = &
Tr [\mu b^{\dagger}_{\vec k} (t) b_{\vec k}(t)]
\nonumber \\
& = & \langle  N_{\vec k}(t_0)\rangle + \mid \beta_{\vec k}(t) \mid^2
[1 +2 \langle  N_{\vec k}(t_0)\rangle],
\end{eqnarray}     
where $\langle  N_{\vec k}(t_0)\rangle=Tr[\mu a^\dagger_{\vec k}a_{\vec k}]$.
 For fermions it decreases the initial number.

The above result can be understood in the parametric oscillator
description as the amplification by a factor ${\cal A} _{\vec k} = 2s_{\vec k}
+ 1$, of a) the vacuum fluctuation, yielding
$|\beta_{\vec k}(t)|^2$, and of b) the particles already present
$ N_{\vec k}(t_0)$, i. e.,
\begin{equation}
( N_{\vec k}(t) )= \mid \beta_{\vec k}(t) \mid^2 + {\cal A}_{\vec k}
\langle  N_{\vec k}(t_0)\rangle
\end{equation}
where $ s_k = |\beta _{\vec k}(t)|^2$.
The second part is called stimulated production. It yields an
energy density $\rho_n$  given by
\begin{eqnarray}
\rho_n & = & \langle n \mid \Lambda^0_0 \mid n \rangle \nonumber \\
  & = & \frac{1}{a^4} \int \frac{d^3k}{(2\pi)^3} ( \mid \chi^\prime_{\vec k}
\mid^2 + \Omega_{\vec k}^2 \mid \chi_{\vec k} \mid^2)\langle
a^\dagger_{\vec k} a_{\vec k}\rangle, \nn \\
  & = & \frac{1}{a^4} \int \frac{d^3k}{(2 \pi)^3} (2s_{\vec k} + 1)
           \Omega_{\vec k} \langle  N_{\vec k}(t_0)\rangle
\end{eqnarray}   
where $\langle  a^\dagger_{\vec k} a_{\vec k}\rangle \equiv \langle
N_{\vec k}(t_0)\rangle= Tr[ \mu a^\dagger_{\vec k} a_{\vec k}]$.
Combining (18) and (23), for a density matrix diagonal in the number state,
the total energy density of particles created from the vacuum and from
those already present in the $n$-particle state is given by
$$
\rho (t) = \rho_0 + \rho_n =
   \frac{1}{a^4} \int \frac{d^3k}{(2\pi)^3} ( \mid \chi^\prime_{\vec k}
   \mid^2 + \Omega_{\vec k}^2 \mid \chi_{\vec k} \mid^2)
   (\frac{1}{2} + \langle a^\dagger_{\vec k} a_{\vec k}\rangle) \nn \\
$$
\begin{equation}
 = \frac{1}{a^4} \int
\frac{d^3k}{(2\pi)^3} {\cal A}_{\vec k} \Omega_{\vec k} (\frac{1}{2} +
 \langle  N_{\vec k}(t_0)\rangle).
\end{equation}
This can be understood as the result of parametric amplification
by the factor $\cal{A}$  of the energy density of vacuum fluctuations
$ \hbar \O/2$ and that of the particles originally present in the $k$th mode
at $t_0$, i.e., $\langle  N_{\vec k}(t_0)\rangle\hbar \Omega_{\vec k}$.

For the special but important case where $\mu$ is thermal at
temperature $T = \beta^{-1}, \; \langle  N_{\vec k}\rangle$ obeys
the Bose-Einstein distribution function for scalar fields.
The magnification of the
$n$-particle thermal state gives the finite-temperature contribution
of particle creation, with energy density
\begin{equation}
\rho_T = \frac{1}{a^4} \int \frac{d^3k}{(2\pi)^3}
(2s_{\vec k}+1)\Omega_{\vec k}/(e^{\beta \Omega_{\vec k}}-1).
\end{equation}
For a massless conformal field, this yields the familiar
Stefan-Boltzmann relation
\begin{equation}
\rho_T = \frac{\pi^2}{30} T^4.
\end{equation}
Finite-temperature particle creation and the related entropy
generation problem have been discussed in \cite{Hu82,VacVis}.

For a more general density matrix the behavior of the induced or stimulated
part of particle creation could increase or decrease,
depending on the correlation and phase relation of the initial state,
even though  the spontaneous creation part always give an increase in particle
number. Both are important factors in the consideration of entropy
generation processes \cite{HuKan}.

\subsection{Evolutionary Operator, Squeezing and Rotation}
\setcounter{equation}{0}
An equivalent description of particle creation is by means of the evolutionary
operator $U$ defined by
\be
b_{\pm \vec k}(t) = U(t) a_{\pm \vec k}U^\dagger (t)
\te
where $UU^\dagger =1$. The form of $U$ was deduced by Parker \cite{Par69}
following Kamufuchi and Umezawa \cite{KamUme}.  In the modern language
of squeezed states \cite{sqst}, one can write $U=RS$ as
a product of two unitary operators, the {\bf rotation operator}
\be
R(\q)=\exp [-i\q (a_+^\dagger a_+ + a_{\_}^\dagger a_{\_})]
\te
and the {\bf two mode squeeze operator}
\be
S_2(r, \f)=\exp [r(a_+a_{\_}e^{-2i\phi}-a_+ ^\dagger a_-^\dagger e^{2i\phi})]
\te
where $r$ is the squeeze parameter with  range $0 \le r <  \infty$ and
$\phi, \q$ are the rotation parameters with ranges $ -\pi/2 <  \phi \le
\pi/2, ~~ 0 \le \q < 2 \pi$. (These parameters and $U, R, S$
should all carry the label $\vec k$.
The $\pm$ on $a$ refer to the $\pm {\vec k}$ modes.) Note that
\be
S^{\dagger}_2 (r, \phi) = S^{-1}_2 (r, \phi) = S_2 (r, \phi + \pi/2)
\te
The three real functions $(\theta_{\vec k}, \phi_{\vec k}, r_{\vec k})$
are related to the
two complex functions $(\a_{\vec k},\b_{\vec k}) $ by
\be
\a_{\vec k} = e^{i\theta_{\vec k}}\cosh r_{\vec k},\;\;
\b_{\vec k} = e^{i(\theta_{\vec k}-2\phi_{\vec k})} \sinh r_{\vec k}.
\te
For mode-decompositions in spatially-homogeneous spacetimes leading
to no mode-couplings, the Bogolubov transformation connecting
the $a_{\vec k}$ and the $b_{\vec k}$ operators is given by
(for more general situations, see \cite{Hu72}):
\be
b_{\pm \vec k} = \a_{\vec k}a_{\pm \vec k}+\b_{\vec k}^* a_{\mp \vec k}
^\dagger.
\te
We see that because of the linear dependence of $b_{+ \vec k}$ on
$a_{+\vec k}$ and $a^{\dagger}_{- \vec k}$ (but not $a^{\dagger}_{+ \vec k}$)
a two-mode squeeze operator is needed to describe particle pairs in states
$\pm {\vec k}$.

The physical meaning of `rotation' and `squeezing' can be seen from the
result of applying these operators for a single-mode harmonic
oscillator as follows: ($\vec k$th  mode label is omitted below)\\

The Hamiltonian is
\be
H_0 = \omega (a^\dagger a + \ha)
\te
Under rotation,
\be
R |0\rangle = |0\rangle, ~~
R a R^ \dagger  = e^{i \theta} a.
\te
Also,
\be
R(\theta) R(\theta') = R (\theta + \theta').
\te
This implies that
\bea
R \hx R^\dagger &=& \cos \q \hx - \sin \q \hp    \nn \\
R \hp R^\dagger &=& \sin \q \hx + \cos \q \hp.
\tea
where
\be
a= \frac{1}{\sqrt 2}(\sqrt \omega \hat x + i \frac{\hat p}{\sqrt \omega}).
\te
Thus the name rotation.
Let $ \D a = a - \langle  a\rangle$, (where $\langle  \rangle$ denotes
the expectation value with respect to any state) then the second-order
`noise moments' of $a$ are defined as \cite{sqst}:
\bea
\langle  (\D a)^2\rangle &=& \langle  a^2\rangle - \langle  a\rangle^2
= \langle (\D a^\dagger)^2 \rangle ^* \nn \\
            &=& \ha [ \langle   (\D x)^2\rangle - \langle  (\D p)^2\rangle] +
i \langle  (\D x \D p)_{sym}\rangle \nn \\
\langle  |\D a|^2\rangle &=& \ha \langle  \D a \D a ^\dagger + \D a^\dagger \D
 a\rangle = \ha [\langle  (\D x)^2\rangle + \langle   (\D p)^2\rangle].
\tea
The first quantity is the variance of $a$, a complex second moment,
while the second is the correlation, a real second  moment, which, as seen
in the more familiar $x, p$ representation, measures the mean-square
uncertainty (called total noise in \cite{sqst}).
Rotation preserves the number operator
\be
R a^\dagger a R^\dagger = a^\dagger a.
\te
It rotates the moment
\be
\langle  R (\D a)^2 R^\dagger\rangle = e^{2i\q} \langle  (\D a)^2\rangle
\te
corresponding to a redistribution between $\hx, \hp$,
but preserves the uncertainty
\be
\langle   R | \D a|^2  R^ \dagger \rangle = \langle   |\D a|^2\rangle.
\te

One can define a {\bf displacement operator} as
\be
D(\m) =\exp[\m a ^\dagger  -\m^* a].
\te
Note that $D^{-1} (\m) = D^{\dagger}(\m) = D(-\m)$.
The coherent state  can be defined as
\be
|\m\rangle=D(\m)|0\rangle.
\te
Thus
\be
a |\m\rangle = \m |\m\rangle,
\te
and
\be
Da^\dagger a D^\dagger = a^\dagger a - (\m a^ \dagger + \m ^* a) + |\m|^2.
\te
Under displacement,
\be
D(\m)a D^\dagger (\m) = a- \m.
\te
The displacement operation also preserves the uncertainty
\be
\langle  D |\D a|^2 D^\dagger \rangle = \langle   | \D a|^2\rangle.
\te

The {\bf single-mode squeeze operator} is defined as
\be
S_1 (r, \f)=\exp \left[\frac{r}{2}(a^2 e^{-2i\phi}-
                                   a^{\dagger 2} e^{2i\phi})\right].
\te
A squeezed state is formed by squeezing a coherent state,
\be
|\s\rangle_{\m} = S_1 (r, \phi) |\m\rangle
\te
or,
\be
|\s\rangle_{\m} = |r,\phi,\m\rangle=S_1 (r,\f)D(\m)|0\rangle.
\te
Call $b= S_1 a S_1^\dagger$, then
\be
b|\s\rangle = \m |\s\rangle
\te
and
\be
b= S_1 a S_1^\dagger = a\cosh r  + e^{2i \f} a^\dagger \sinh r.
\te
Thus a squeezed state in the Fock space of $a$ becomes a coherent state
in the Fock space of $b$ with the same eigenvalue.
 From this we see the result of $S_1$ acting on  $\hat x$ and $\hat p$ :
\bea
S_1 \hx S_1 ^\dagger  &=&(\cosh r+\cos 2\phi\sinh r)\hx
+(\sin 2\phi\sinh r) \hp \nn \\
S_1 \hp S_1 ^\dagger  &=&(\cosh r-\cos 2\phi\sinh r)\hp+
(\sin 2\phi\sinh r) \hx.
\tea
For $\f =\p/2$, these give
\be
S_1 \hx S_1 ^\dagger =e^{-r}\hx,\;\;\;S_1 \hp S_1 ^\dagger =e^r \hp.
\te
Hence the name `squeezing'. Two successive squeezes, with the same rotation
parameter, result in one squeeze with the squeeze parameter as the sum of the
two parameters:
\be
S_1(r, \f) S_1(r', \f) = S_1 (r+r', \f).
\te
The expectation value of squeezing the number operator is
\be
\langle  S_1^\dagger a^\dagger a S_1\rangle = \sinh^2 r + (1+2 \sinh^2 r)
\langle  a^\dagger a\rangle
                               + \sinh 2r Re [ e^{-2i\f} \langle  a^2\rangle]
\te
and that of the correlation is
\be
\langle  S_1^\dagger |\D a|^2 S_1\rangle = \cosh 2r \langle  |\D a|^2\rangle
+ \sinh 2r Re [ e^{-2i\f} \langle  (\D a)^2\rangle]
\te
which for the vacuum and coherent states is always greater than or equal to the
original value.

The two-mode squeeze operator defined before
is more suitable for the description of cosmological particle creation.
One can show that the $out$ state is generated from the $in$ state by
including contributions from all $k$ modes,
\be
|out\rangle = RS |in\rangle, ~~ {\rm or}~~ |) = RS | \rangle
\te
where
\be
S= \Pi^\infty_{\vec k =0} S_2 (r_{\vec k}, \phi_{\vec k})
\te
In general
\be
\langle  out| F(b_\pm, b_\pm^\dagger ) |out\rangle=
\langle  in | F (a_\pm, a_\pm^\dagger)|in\rangle.
\te
The $|in\rangle$ state can be a number state, a coherent state or a squeezed
 state.
If the initial state  is a vacuum state, $|in\rangle= |0 in\rangle$, then
\be
|0 out\rangle= S (r, \f -\q) |0 in\rangle
\te
where
\be
S(r, \f-\q)= exp \{ \Sigma_{\vec k} r_{\vec k} [ e^{-2i(\f_{\vec k}-\q_{\vec k}
)}
a_{\vec k} a_{-\vec k} - e^{ 2i(\f_{\vec k}-\q_{\vec k})} a_{\vec k}^\dagger
a_{-\vec k}^\dagger ] \}
\te
The squeeze parameter  $\sinh^2 r_{\vec k} = |\b_{\vec k}|^2 $ measures
the number of particles created. Rotation does not play a role.
Thus, as observed by Grishchuk and Sidorov \cite{GriSid},
cosmological particle creation amounts to squeezing the vacuum.
The same can be said about Hawking radiation \cite{Haw75}.
For a massless scalar field in an eternal black hole,
call $e^{iJ}$ the unitary operator which connects the Kruskal vacuum with
the Schwarzschild vacuum (see e.g., \cite{BirDav})
\be
|0\rangle_S = e ^{iJ} |0\rangle_K
\te
where
\be
iJ= \Sigma_{\vec k} \tanh^{-1}(e^{-\pi \omega / \kappa})
(b_{-\vec k}^{(1)} b_{\vec k}^{(2)} -
 b_{-\vec k}^{(1)\dagger} b_{\vec k}^{(2)\dagger}).
\te
Then the squeeze and rotation parameters can be identified as
\be
r_{\vec k} = \tanh^{-1}( e^{-\p \o / \k}), ~~ \phi_{\vec k} = \q _{\vec k},
\te
where $\k$ is the surface gravity of the black hole. This is the well-known
expression for Hawking radiation.
We see that for low-momentum modes in
a black hole of high temperature, the squeezing is strong.

\newpage
\section{Number, Coherence and Initial States}
\subsection{Number does not always increase}
\setcounter{equation}{0}
We will show in this section that the number of particles produced
depends very much
on the initial state chosen. The common impression of a net number increase
associated with cosmological particle creation is premised upon
the assumption that the initial state is an eigenstate of the  number
operator (called `number state' for short here), and an implicitly invoked
random-phase approximation. For states other than this, or for fermions,
this is not necessarily true. This was already pointed out in
\cite{HuPav,Kan,HuKan}.
We shall show this explicitly for the coherent state and the squeezed states.

The number operator for a particle pair in mode $k$ is given by
\be
N=a_+^{\dagger} a_+ + a_{\_}^{\dagger} a_{\_}.
\te
The expectation value of the number operator with respect to the $|out\rangle$
vacuum for a general initial state is
\begin{eqnarray}
(N)= \langle S_2^{\dagger} R^{\dagger} N RS_2\rangle & = &
2|\b|^2 +(1+2|\b|^2)\langle  N\rangle \nonumber \\
 & - & 2|\a||\b|(e^{2i\phi}\langle  a_+^\dagger a_{\_}^\dagger \rangle
       +e^{-2i\phi}\langle a_+a_{\_}\rangle).
\end{eqnarray}
Comparing this expression with (2.22), the factor of two for the first
$|\b|^2 $ term comes from the spontaneous creation of particles in the
$\pm \vec k$ modes.
The net change in the particle number from the initial to the final
state is
\be
\D N \equiv (N)- \langle N \rangle =
2|\b|^2[1+\langle  N\rangle]-2|\b||\a| \{e^{2i\phi}\langle  a_+^{\dagger}
a_{\_}^{\dagger}\rangle+e^{-2i\phi}\langle  a_+a_{\_}\rangle\}.
\te

Here, the first two terms in the square brackets are respectively
the spontaneous and stimulated emissions
and the last term in the curly brackets is the interference term.
The difference between spontaneous and stimulated creation of particles
in cosmology was explained first by Parker \cite{Par69} and
explored in more detail by Hu and Kandrup \cite{HuKan}.
Note that since there is no $\theta$ dependence, rotation has no effect.
If $r_{\vec k} \ne 0$ for some $\vec k$
both spontaneous and stimulated contributions are positive.
The interference term can be negative for states which give nonzero
$\langle   a_+ a_{-}\rangle $.
Only when this term is non-zero can $\Delta N$ be negative.

We will calculate the change in particle number for some specific initial
states.

\noindent {\bf a. number state}

For an initial number state $ |n \rangle = |n_+,n_{\_}\rangle$
\be
\D N=2|\b|^2(1+n_+ +n_{\_}).
\te
We see that the number of particles will always increase.

\noindent {\bf b. coherent state}

For an initial coherent state
\be
|\mu\rangle=D(\mu_+)D(\mu_{\_})|0,0\rangle
\te
we find that
\be
\D N=2|\b|^2[1+\langle  N_+\rangle+\langle N_{\_}\rangle]
-4|\b||\a|\sqrt{\langle  N_+\rangle\langle  N_{\_}\rangle}
\cos (2\phi-\zeta_+-\zeta_{\_}),
\te
where
\be
\mu_+=\sqrt{\langle  N_+\rangle} e^{i\zeta_+},\;\;\;
\mu_{\_}=\sqrt{\langle  N_{\_}\rangle}e^{i\zeta_{\_}}.
\te
Note the existence of the interference term which can give a negative
contribution. It depends not only on the squeeze parameters $|\b|$ and
$\f$, but also on the particles present and the phase of the initial
coherent state.
Conditions favorable to a decrease in $\Delta N$ are
$\cos (2\phi-\zeta_+-\zeta_{\_})=1$ and
$\langle  N_+\rangle=\langle  N_{\_}\rangle=\langle  N\rangle/2$.
In this case we find $\Delta N$ is negative if
\be
\langle  N\rangle >\frac{|\b|}{|\a|-|\b|}.
\te
{\bf c. single-mode squeezed vacuum state}

For an initial one-mode squeezed state
\be
|\sigma\rangle_1=S_{1+}(r_+,\phi_+)S_{1-}(r_-,\phi_-)|0,0\rangle
\te
generated by squeezing the vacuum with $S_{1\pm}$ for the $\pm \vec k$ modes,
we get
\be
\D N = 2|\b|^2(1+\langle  N_+\rangle+\langle  N_{\_}\rangle).
\te
Once again particle number will always increase.

\noindent {\bf d. two-mode squeezed vacuum state}

For an initial two-mode squeezed vacuum
\be
|\sigma\rangle_2=S_2(r_0,\phi_0)|0,0\rangle
\te
where $S_2$ is defined earlier,
\be
\D N = 2|\b|^2[1+\langle  N\rangle]+
2|\b||\a|\sqrt{\langle  N\rangle(2+\langle  N\rangle)}\cos 2(\phi-\phi_0).
\te
The cosine factor shows that particle number can decrease given the right
phase relations.
It can be shown that for $\cos 2(\phi-\phi_0)=-1$
particle number would decrease
($\D N \le 0$) if  $r_0 \ge  r/2$.
If the phase information is randomized the cosine factor averages to zero and
there is a net increase in particle number.
Since a squeezed state is the end result of squeezing a vacuum via
particle creation, one might naively expect to see a monotonic increase in
number. Our result shows that this is true only if the phase information is
lost in the squeezed state to begin with.

In summary we can make the following observations:

\noindent a) Rotation $R$ in the evolution operator $U=RS$ does not influence
particle creation.

\noindent b) For an initial number state or single mode squeezed vacuum we find
a net increase in the number of particles.

\noindent c) For an initial coherent state and two mode squeezed vacuum,
particle number
can increase or decrease. A net increase can nevertheless be obtained
by suitable choices of $S_2(r, \f)$ and $S_2(r_0, \f_0)$.

\noindent d) If random phase is assumed for the initial state
the interference term can be averaged out to zero and there will be
a net increase in number of particles.

\subsection{Coherence can persist}

A measure of the coherence of the system is given by the uncertainty
(called variance in \cite{Hu72}, \cite{Mollow} and \cite{HuPav})
\be
|\D a|^2 =\frac{1}{2}(\D a\D a ^\dagger  + \D a ^\dagger  \D a)
\te
where $\D a=a-\langle  a\rangle$.
The expectation value of the uncertainty with respect to a state $|\psi\rangle$
is thus,
\be
\langle  \psi||\D a|^2|\psi\rangle=\langle  \psi|a ^\dagger  a|\psi\rangle-
|\langle  \psi|a|\psi\rangle|^2 + \frac{1}{2}.
\te
The expectation value of the uncertainty with respect to a tranformed state
$ |\psi) \equiv RS |\psi\rangle$ is given by
\be
(\psi| |\D a|^2 |\psi) = \cosh 2r \langle  \psi||\D a|^2|\psi \rangle
-2 \sinh 2r Re [ e^{-2i\f} \langle  \psi | \D a_+ \D a_- |\psi\rangle]
\te
where $|\D a|^2 \equiv |\D a_+|^2 + |\D a_-|^2$.
For an initial number state, $|\psi\rangle= |n\rangle$,
\be
(n| |\D a|^2 |n)= 2(\frac{1}{2}+ |\b|^2) \langle n| |\D a|^2 |n\rangle~
                          \ge ~ \langle n| |\D a|^2 |n\rangle
\te

For a coherent state, $|\psi\rangle = |\m\rangle$
\be
(\m| |\D a|^2 |\m) = 2 (\frac{1}{2} + |\b|^2)
\langle  \m| |\D a|^2 |\m\rangle
{}~ \ge ~ \langle  \m| |\D a|^2 |\m\rangle
\te
where the first term corresponds to the vacuum fluctuation and the second
term [whose sum over all modes is equivalent to
$ tr(v ^\dagger _{\vec k} v_{\vec k})$ in \cite{Hu72,HuPav}]
measures the mixing of the positive and negative frequency components of
different modes. This result was first derived in \cite{Hu72},
and discussed further in \cite{HuPav}. Notice that it is always greater
than the original value $ \langle  |\D a|^2\rangle_\m$.

For a squeezed state,
$|\psi \rangle = |\s \rangle = S_2 (r_0, \f_0) |\m\rangle $
\be
(\s| |\D a|^2 |\s) = \cosh 2r \langle  \s||\D a|^2 |\s \rangle
 -2 \sinh 2r Re [ e^{-2i\f} \langle   \s| \D a_+ \D a_- |\s \rangle] ,
\te
which can be smaller than the initial value.

Notice that of the three states we discussed, only the squeezed state can
allow for a decrease in the uncertainty, i.e., an increase in the coherence
as the system evolves.
In addition, even though the total number and the total uncertainty of the
initial state of the two modes change with particle creation, their difference
remains a constant. This is because cosmological particle creation is
described by the two mode squeezed operator which satisfies the
relations:
$$
  \langle \psi |S^\dagger (a^\dagger_+ a_+ -a^\dagger _- a_-) S |\psi \rangle~
=~\langle \psi            |a^\dagger_+ a_+ -a^\dagger _- a_-    |\psi \rangle~,
$$
\be
  \langle \psi |S^\dagger (|\D a_+|^2 - |\D a_-|^2) S |\psi \rangle~
=~\langle \psi |          (|\D a_+|^2 - |\D a_-|^2)   |\psi \rangle~.
\te
\newpage
\section{Fluctuations in Number}
\setcounter{equation}{0}

Spontaneous particle creation can be viewed as the parametric
amplification of vacuum fluctuations (or squeezing the  vacuum).
Particle number
is an interesting quantity as it measures the degree the vacuum is
excited. The fluctuation in particle number is another
interesting quantity, as it can be related to the noise of the quantum
field and the susceptibility of the vacuum.
This is similar in nature to the energy fluctuation (measured by the heat
capacity at constant volume) of a  system
being related to the thermodynamic stability of a canonical
system, or the number fluctuation (measured by
the compressibility at constant pressure) of a system
being related to the thermodynamic stability of a grand canonical system.
In gravity, we know that the number fluctuation of a self-gravitating system
can be used as a measure of its heat capacity (negative) \cite{LynBel};
and those associated with particle creation from a black hole can be used
in a linear-response theory description
as a measure of the susceptibility of spacetime \cite{CanSci,Mot}.
We expect that this quantity associated with cosmological
particle creation may provide some important information about
quantum noise and vacuum instability \cite{nfsg,HMLA}.

Define $\d_i O \equiv [\langle  O^2\rangle- \langle  O\rangle^2]$
as the variance or mean-square fluctuations of
the variable $O$ with respect to the initial state $|~\rangle$,
and the corresponding quantity $\d_f O$ as that with respect to the
final state $|~)$. Consider the difference between the final and
the initial number fluctuation of both the $\pm$ kinds,
\be
\d N = (\d_f N_+ + \d_f N_-) - (\d_i N_+ + \d_i N_-).
\te
Using the expressions given in Sec. 2, we obtain
\bea
\d N & = & 2 |\a|^2|\b|^2 [ \d N_+ + \d N_- + \d L + \partial (N_+ N_-)]_i\nn\\
     &   & -(|\a|^3 |\b| + |\a| |\b|^3)[\partial (N_+L) + \partial (N_-L)]_i
\tea
where the subscript $i$ refers to the fact that the expectation values
are taken with respect to the initial states $|~\rangle$, the symbol $\partial$
denotes
\be
\partial (PQ) \equiv [\langle  PQ\rangle + \langle  QP\rangle -2 \langle
 P\rangle\langle  Q\rangle]
\te
and
\be
L=e^{2i\phi}a_+ ^\dagger a_- ^\dagger  + e^{-2i\phi}a_-a_+.
\te

Now for an initial number state $|n \rangle = |n_+, n_-\rangle$,
\be
\d N = 2 |\a|^2|\b|^2(1+ n_+ +n_{\_} +2n_+ n_{\_}).
\te
we see that the number fluctuations will always increase.
For an initial coherent state
$|\mu\rangle = D (\mu_+) D (\mu_{\_}) |0, 0\rangle$,
where $ \m_{\pm} = \sqrt{\langle N_{\pm}\rangle} e^{i \zeta_{\pm}} $,
\begin{eqnarray}
\d N & = & 2|\a|^2|\b|^2[1+ 2(\langle  N_+\rangle+\langle  N_{\_}\rangle)]
\nonumber \\
 & - & 4\sqrt{\langle  N_+\rangle\langle  N_{\_}\rangle}
(|\a|^3 |\b| + |\a| |\b|^3)\cos(2\phi-\zeta_+ -\zeta_{\_}).
\end{eqnarray}
We find that under the conditions $\cos(2\phi-\zeta_+ -\zeta_{\_}) =1$
and $\langle N_+\rangle=\langle N_-\rangle=\langle N\rangle/2 $
\be
\langle N \rangle > \frac{|\b||\a|}{|\a|^2+|\b|^2-|\b||\a|}
\te
$\d N$ can be negative. In the weak particle creation limit $|\b|\rightarrow
 0, |\a|\rightarrow 1$ we find that (4.7) is equivalent to (3.8).
Clearly conditions for a decrease in number fluctuations
are not the same as those for a decrease in the number.

For a single-mode squeezed state $|\sigma\rangle_1 = S_{1+} (r_+,\f_+)
S_{1-} (r_-,\f_- ) |0, 0\rangle$
\bea
\d N & = & 2|\a|^2|\b|^2[(1+\langle N_+\rangle+\langle N_-\rangle)^2
+ \langle N_+\rangle (1+ \langle N_+\rangle)
+ \langle N_-\rangle (1+ \langle N_-\rangle)               \nonumber \\
& - & 2\sqrt{\langle N_+\rangle (1+\langle N_+\rangle)
             \langle N_-\rangle (1+\langle N_-\rangle)}
        \cos 2(2\phi-\phi_+-\phi_-)].
\tea
 From this it can be shown that, like the change in number, the change in the
number fluctuations will always be positive for an initial single mode squeezed
vacuum.

For a two-mode squeezed state
$ |\sigma\rangle_2 = S_2 (r_0,\f_0)|0, 0\rangle$
\bea
\d N & = & |\a|^2|\b|^2\{2(1+\langle N\rangle)^2
+\langle N \rangle (2+ \langle N\rangle) [ 1+ \cos 4(\phi-\phi_0)]\} \nn \\
& + & 2(|\a|^3|\b|+|\b|^3|\a|)(1+\langle N\rangle)
\sqrt{\langle N\rangle(2+ \langle N\rangle)}\cos 2(\phi-\phi_0).
\tea
Note that there is  no definite relation between $\D N$ and $\d N$.
For large $N>>1$ or small $|\b|<<1$, $\d N \le 0$.
The relevance of the number fluctuations in cosmological particle creation
in defining the susceptibility of the vacuum and the noise of
quantum fields have been hinted upon earlier \cite{VacVis,GraEnt,HuPhysica}.
The result obtained here will be useful in relating to issues
of noise and fluctuation of quantum fields, and dissipation and instability
of spacetime in semiclassical gravity and quantum cosmology
\cite{fdrc,nfsg,HMLA,HuMat3}.

\vskip 1cm
{\bf Acknowledgement} This work is supported in part by the National
Science Foundation under grant PHY91-19726.


\end{document}